# Energy-tunable sources of entangled photons: a viable concept for solid-state-based quantum relays

Rinaldo Trotta,[1]* Javier Martín-Sánchez,[1] Istvan Daruka,[1] Carmine Ortix,[2] ** and

Armando Rastelli[1]

[1]Institute of Semiconductor and Solid State Physics, Johannes Kepler University Linz,

Altenbergerstr. 69, A-4040 Linz, Austria.

[2]Institute for Theoretical Solid State Physics, IFW Dresden, Helmholtzstr. 20, D-01069

Dresden, Germany.

We propose a new method of generating triggered entangled photon pairs with wavelength on demand. The method uses a micro-structured semiconductor-piezoelectric device capable of dynamically reshaping the electronic properties of self-assembled quantum dots (QDs) via anisotropic strain-engineering. Theoretical models based on **k·p** theory in combination with finite-element calculations show that the energy of the polarization-entangled photons emitted by QDs can be tuned in a range larger than 100 meV without affecting the degree of entanglement of the quantum source. These results pave the way towards the deterministic implementation of QD entanglement resources in all-electrically-controlled solid-state-based quantum relays.

**Corresponding Author***

Rinaldo Trotta

Institute of Semiconductor and Solid State Physics,

Johannes Kepler University Linz, Altenbergerstr. 69,

A-4040 Linz, Austria.

Tel.: +43 732 2468 9599Fax: +43 732 2468 8650

e-mail: rinaldo.trotta@jku.at

**e-mail: c.ortix@ifw-dresden.de

Quantum communication deals with the transfer of quantum states from one place to another. One of its applications, quantum key distribution, promises secure data transmission by encoding "bits", e.g., in the polarization state of photons[1]. However, unavoidable losses in the transmission channel limits the reachable distances to a few hundred of km. One elegant solution to this problem is the use of quantum repeaters[2], which mitigate losses by means of *quantum relays* (QRs), *i.e.*, devices capable to teleport, or "swap" entanglement between distant nodes. A single QR, sketched in Figure 1a, consists of two entanglement resources (ERs) emitting entangled-photon-pairs[3]. A Bell-state measurement (BSM) between photons stemming from two remote ERs allows entanglement to be distributed over distant nodes. A partial BSM can be obtained by Hong-Ou-Mandel (HOM) type two-photon interference,[4] already employed in the first demonstration of entanglement swapping[5], in quantum teleportation experiments in a QR configuration[6,7], and recently implemented using single photons from independent sources[8,9], including semiconductor quantum dots (QDs)[10,11].

QDs are semiconductor nanostructures that can be easily integrated into well-established devices[12] and are capable of deterministic generation of high-quality single[13] and entangled[14] photons. In particular, polarization-entangled photon-pairs can be generated during the cascade of a biexciton (XX) to the exciton (X) to the ground state[15], (Figure 1b). With such a scheme in mind, one could dream to build-up a QR with two identical QD-ERs and to use HOM interference[4] of the XX photons to entangle the X photons. However, this simple picture requires a number of extraordinary challenges to be overcome. Problems related to the need of Fourier-limited[16,17], bright[18,19,20] and site-controlled[21] photon-sources have been successfully addressed. There is another issue, however, which is still open: different from real atoms, each QD possesses its own size, shape, and composition[22] and, as a consequence, unique emission spectra. This has a dramatic impact on the proposed scheme for QRs (see Figure 1c). First, there are no identical QDs in reality. Considering that the inhomogeneous broadening of QD emission is ~ tens of meV, the probability of finding two QDs with XX (or X) emission at the same energy within typical radiative-limited linewidth (~µeV) is $<10^{-4}$. This eventually hinders BSM, because the colour difference between the photons prevents their interaction at the beam splitter[10,11]. Second, the anisotropic electron-hole exchange interaction[23] – which generally leads to the appearance of a fine-structure-splitting (FSS, $s$) between the two bright X states (see Fig. 1c) – limits the capability of a QD to generate photon-pairs featuring high entanglement degree[12,24,25]. Recent calculations[26] show that a very low portion of as-grown QDs are free of asymmetries (~$10^{-2}$) and the numbers increase slightly if sophisticated growth protocols are employed[21]. Therefore, the probability of finding two as-grown QDs for a QR is ~$10^{-9}$ or less. These hurdles have naturally led to the idea of post-growth tuning of the QD emission properties via the application of external perturbations, such as strain[27], electric[28], and magnetic[29,30] fields. It turned out, however, that even with the aid of external "knobs" it is extremely difficult to implement the scheme of Figure 1b, since any attempt to control the energy of the entangled photons from remote QDs generally restores the FSS and spoils entanglement. In summary, *single QDs* are very promising ERs for applications, as recently demonstrated by quantum teleportation experiments performed using QDs in entangled-light-emitting diodes (ELEDs)[31]. However, a quantum network based on QRs using *several QDs* is at present out of reach.

In this Letter, we propose a viable concept for an energy-tunable source of polarization-entangled photons, which can be exploited for the implementation of scalable QRs. The key idea, backed by theoretical models for the exciton Hamiltonian, is to use anisotropic-strain engineering of the semiconductor matrix hosting the QDs to reshape their electronic structure so that entanglement swapping becomes possible. We then propose a device (see Figure 1d) that can successfully address this task: It consists of a micro-machined piezoelectric-actuator featuring six legs and capable of deforming *in any direction* a nanomembrane containing QDs. Finite-element methods (FEM) combined with **k·p** models are used to demonstrate that (*i*) the application of three voltages on pairs of legs allows stress anisotropies to be engineered on demand with magnitudes as large as 3.6 GPa and (*ii*) the energy of the entangled photons emitted by QDs can be modified over a spectral range >100 meV. Furthermore, the use of diode-like nanomembranes[32] featuring additional electrical contacts opens up the unprecedented possibility to inject carriers electrically into the QDs, thus leading to wavelength-tunable ELEDs. Different from other approaches demonstrating QRs with independent sources[33,34], the QR scheme we propose here makes use of deterministic – both in time and wavelength – and electrically-controlled ERs and it is, at least in principle, scalable.

We begin presenting the theory underlying the development of energy-tunable sources of entangled photons. In standard QDs, the coupling of electrons and heavy-holes results in two bright excitonic states. The symmetry of the confining potential is of fundamental importance for the fine structure of the bright doublet[23]. A symmetry lowering from $C_{4v}/D_{2d}$ to $C_{2v}$ leads to two states $|B_\pm\rangle = |1\rangle \pm |-1\rangle$ split by the FSS. In real QDs, strain, piezoelectricity, and alloying further lower the symmetry to $C_1$. Besides a

renormalization of the FSS, these effects introduce an additional mixing of the $|B_\pm\rangle$ states and cause the polarization direction of the exciton emission ($\theta_\pm$) to depart from the [110] and [1-10] crystal directions generally expected for $C_{2v}$ QDs[35]. To show this, we consider an effective two-level Hamiltonian in the $|B_\pm\rangle$ basis:

$$H = E_0\tau_0 + \eta\tau_z + k\tau_x, \qquad (1)$$

where $\tau$'s are the Pauli matrices with $\tau_0 = I_2$, *i.e.*, the identity matrix. $E_0$ is the energy of an exciton confined in a $C_{4v}/D_{2d}$ QD, whereas $\eta$ and $k$ account for the lowering of the symmetry to $C_1$ and lead to two non-degenerate transitions at $E_\pm = E_0 \pm \sqrt{\eta^2 + k^2}$. $\theta_\pm$ can be inferred from the corresponding eigenvectors, $|\psi_\pm\rangle = k|B_+\rangle + \left(-\eta \pm \sqrt{\eta^2 + k^2}\right)|B_-\rangle$, implying $\tan\theta_\pm = \left(-\eta \pm \sqrt{\eta^2 + k^2}\right)/k$. We point out that $|B_+\rangle$ ($|B_-\rangle$) is aligned along the [110] ([1-10]) direction. Because the two excitons now belong to $\Gamma_1$, the Wigner-von Neumann non-crossing rule[36] is in full force and single external fields are in general not sufficient to suppress the FSS[35,37]. As shown in previous works[24,38], this problem can be circumvented in *any QD* by the simultaneous application of *two* external fields. This is possible because they are *always* capable of satisfying the constraints of level degeneracy, that is $\langle B_+|H_{eff}|B_-\rangle \equiv 0$ and $\langle B_+|H_{eff}|B_+\rangle \equiv \langle B_-|H_{eff}|B_-\rangle$, where $H_{eff}$ is the Hamiltonian in the presence of the two fields. At this point, it is intuitive that external fields with *three* independent degrees of freedom are required to build-up an energy-tunable source of entangled photons: *two* are taken up to fulfill the condition of level degeneracy, while the *third* is needed to control the X energy. In-plane stress fields naturally offer *three* independent degrees of freedom, as they are characterized by three components of the stress tensor, $\sigma_{xx}$, $\sigma_{yy}$, $\sigma_{xy}$ or, equivalently, by two principal stresses $S_1$, $S_2$ and an angle $\phi$ with respect to the [110] crystal direction. The strain Hamiltonian can be written as

$$\delta H_s = \bar{\alpha} \cdot \bar{p} \tau_0 + \alpha \cdot \Delta p \cdot \cos(2\phi) \tau_z + \gamma \cdot \Delta p \cdot \sin(2\phi) \tau_x \tag{2}$$

where $\bar{\alpha}$, $\alpha$, $\gamma$ are parameters related to the elastic constants renormalized by the valence-band deformation-potentials, $\bar{p} = S_1 + S_2$ (hydrostatic part), and $\Delta p = S_1 - S_2$ (stress anisotropy). Three parameters are relevant: $\bar{p}$, $\Delta p$, and $\phi$. Their effect on the Hamiltonian is twofold. First, by properly setting $\phi$ such that the *strain* principal direction $\phi_\varepsilon$ is parallel to $\theta_+$, the level degeneracy can be restored by adjusting $\Delta p$. This can be seen by requiring level degeneracy, i.e, $\eta = -\alpha \cdot \Delta p \cdot \cos(2\phi)$ and $k = -\gamma \cdot \Delta p \cdot \sin(2\phi)$ and by noting that $\phi$ is related to $\phi_\varepsilon$ via

$$\tan 2\phi = \frac{S_{44}}{2(S_{11} - S_{12})} \tan 2\phi_\varepsilon,$$

where $S_{ij}$ are the elastic compliance constants. The FSS can now be suppressed once the stress is applied at an angle $\phi = \phi^*$ such that

$$\tan 2\phi_\varepsilon^* = \frac{2(S_{11} - S_{12})}{S_{44}} \frac{\alpha}{\gamma} \tan 2\theta_+,$$

yielding $\phi_\varepsilon^* \approx \theta_+$ when the prefactor is ~1. In order to give a realistic estimate of $\phi_\varepsilon^*$, we account for strain-induced effects using the Dresselhaus-Kip-Kittel Hamiltonian[39] which allows us to calculate $\alpha$ and $\gamma$ for bulk GaAs and InAs. In particular, the following values have been used[39] in the calculation for InAs (GaAs): $\alpha = 0.04$ eV/GPa ($\alpha = 0.033$ eV/GPa) and $\gamma = 0.07$ eV/GPa ($\gamma = 0.039$ eV/GPa). Apart from a slightly larger deviation for the case of GaAs QDs, Figure 2a clearly shows that the excitonic degeneracy can be restored when stress is applied at an angle $\phi^*$ such that $\phi_\varepsilon^* \approx \theta_+$. This can be also seen in Figure 2b, where the FSS is plotted against the X energy-shift for different $\phi_\varepsilon^*$ in an InAs QD featuring $\theta_+ = 16.2°$. The $s=0$ condition can be achieved only for $\phi_\varepsilon^* = 18.5°$. For all other angles, a lower bound of the FSS is observed. We highlight once more that two control-parameters ($\phi$ and $\Delta p$) are required for $s=0$, in line with previous experimental and theoretical results[24,38,38,40,41]. The second strain-related effect comes about the hydrostatic contribution $\bar{p}$ (third control-parameter), which influences the X energy without affecting the FSS. The black lines of Figure 2b show that it is indeed possible to find values of $S_1$ and $S_2$ such that $\phi^*$ and $\Delta p^*$ are constant at $s=0$, while $\bar{p}$, proportional to the X energy, varies. These theoretical calculations also suggest a three-step procedure to achieve this result experimentally: (*step 1*) align the major *stress* axis to $\phi^*$ such that $\phi_\varepsilon^* \approx \theta_+$; (*step 2*) change the magnitude of $\Delta p = S_1 - S_2$ keeping fixed $\phi$ at $\phi^*$ until the condition $s=0$ is reached. This occurs for $\phi = \phi^*$ and $\Delta p = \Delta p^*$. In the Figure 2b, step 2 is performed varying $S_1$ at $S_2 = 0$; (*Step 3*) modify $\bar{p} = S_1 + S_2$ at fixed $\phi = \phi^*$ and $\Delta p = \Delta p^*$ so as to change at will the X energy. In the Figure 2b this is achieved by sweeping again $S_1$ at different values of $S_2$. Since the QD parameters are fixed, the condition of $s=0$ is found exactly at $\phi = \phi^*$ and $\Delta p = \Delta p^*$ but for a different combination of $S_1$ and $S_2$ and, as a consequence, for different X energies. The range of tunability depends on the QD structural details and on the magnitude of $S_1$ and $S_2$ reachable in practice. Using QD parameters estimated in previous works[24,26], assuming maximum values of $S_1$ and $S_2$ ~ 1.8 GPa, and taking into account the experimental value[42] of the shift of the X energy with stress ($\bar{\alpha} = -29$ meV/GPa, so as to include strain-related effects on the conduction band), we predict that at $s=0$ the X energy can be controlled in a range as large as 100 meV (see Figure 2b). It is clear that the success of our proposal is strictly connected to the capability of achieving independent control of $S_1$, $S_2$, and $\phi$, a non-trivial task that we address in the following.

The device concept we propose here as energy-tunable ER consist of a ~300 nm-thick Al(GaAs) nanomembrane containing In(Ga)As QDs connected and suspended on a micro-machined single crystal [Pb(Mg$_{1/3}$Nb$_{2/3}$)O$_3$]$_{0.72}$-[PbTiO$_3$]$_{0.28}$ (PMN-PT) featuring six legs aligned at 60° with each other[43], see sketch of Figure 1d. Quasi-uniaxial stresses are achieved by applying three independent voltages (V$_1$, V$_2$, V$_3$) on contacts defined at the bottom of the legs, while the top is electrically grounded. The voltages induce electric fields in the piezo-legs that lead to an in-plane contraction/extension of the material and, in turn, to a deformation of the nanomembrane in the same direction. The same voltage is applied to opposite legs to limit displacements of the central structure.

To demonstrate that the proposed device allows for full control over $S_1$, $S_2$, and $\phi$, we have performed FEM simulations with realistic parameters for sizes and elastic/piezoelectric properties for PMN-PT[44] and GaAs[45]. In the regime in which both piezoelectric and elastic response are linear, the voltage sets required to obtain a given stress configuration can be predicted using the expression (V$_1$, V$_2$, V$_3$) = $\underline{R}^{-1}$(σ$_{xx}$, σ$_{yy}$, σ$_{xy}$), where $\underline{R}$ is a 3×3 transfer matrix which can be obtained using a small set of FEM simulations. Our FEM analysis suggests that by using reasonable voltages on the piezo-legs we can achieve *full control* of in-plane stress: The top (bottom) panel of Figure 3a shows a FEM simulation for (V$_1$,V$_2$,V$_3$)=(0,600,0)V ((V$_1$,V$_2$,V$_3$)=(600,0,0)V). These voltages allow the $S_1$ axis to be rotated by 120° with respect to the [110] direction, with magnitudes as large as 1.8 GPa. The FEM simulations can now be combined with the $\mathbf{k}\cdot\mathbf{p}$ model to estimate the tunability of X energy at $s=0$. As detailed above, we are interested in tuning the X energy at fixed $\Delta p$ and $\phi$. The results are displayed in Figure 3b, where the X energy-shift is plotted in cylindrical coordinates against $S_1 - S_2$ and $\phi$. Considering that the $\Delta p$ values required to cancel the FSS in QDs[24] are in the MPa range (see the Figure 2b), Figure 3b shows a X energy-shift up to the impressive value of 100 meV. We have therefore demonstrated that the proposed device acts as energy-tunable ER that can be used to build up chains of QRs (see Figure 1e). We point out that the device design enables full optical access to the QDs and it is therefore compatible with two-photon excitation schemes[46] recently employed for on-demand generation of entangled-photons[17]. The built-in metal layer at the bottom of the nanomembrane can act as a back mirror for photonic cavities. Beside the metal-semiconductor-dielectric planar cavity we have implemented in previous works[32], more advanced approaches can be pursued for boosting the flux of QD photons. Among others[47], photonic membranes featuring deterministic GaAs-microlenses[48] are particularly promising, since they can be easily integrated in our device, they would allow achieving extraction efficiency as high as 60%[47,49], they are rather insensitive to strain[32] and electric field and are therefore suitable for broad-band enhancement of XX-X lines. Furthermore, the QDs can be embedded in diode-like nanomembranes[32,25] via gold-thermo-compression bonding. The gold layers sketched in Figure 1d on the top and bottom of the nanomembrane can be exploited as additional contacts for controlling the electric field across the QDs, which can be either used to inject carriers electrically, as in ELEDs[12], or – in synergy with strain - to modify with high speed the QD emission properties. The latter operation mode results in a four-knob device that would allow for independent control of X and XX energies[50] without restoring the FSS. The implementation of ideal QRs (see Fig. 1b) made of identical (in terms of the X and XX energies) QD-ERs would be then feasible. This, in turn, could pave the way towards the realization of a complete quantum repeater[2] featuring quantum memories without heralding[3], since it would allow, *e.g.*, for the storing of the X photons in warm atomic vapors[2] while the XX photons are taken up for partial BSM.

In summary, the implementation of the concept we propose here will lead to a quantum device that fulfils all the points of the "wish-list" of the perfect source of entangled photons[51] with the plus of being tunable in wavelength. Besides quantum networking via QRs, this novel ER could be used for scopes beyond quantum communication as, for example, for fundamental tests of quantum mechanics[52] and for solving quantum computational tasks[53].

**Figure 1. (a)**. Sketch of a QR. Yellow lines represent entanglement. Photodetectors are in black. **(b)** Implementation with two identical QDs using the XX-X-0 cascade. $\sigma^+$ ($\sigma^-$) indicates right (left) circularly-polarized photons. A partial BSM is performed via two-photon interference at a beam splitter followed by polarization-resolved detection. **(c)** Realistic situation for two random QDs. H (V) indicates horizontally (vertically) polarized photons. **(d)**. Sketch of the proposed device. The top and bottom views are depicted on the left and right panel, respectively. Three independent voltages ($V_{1,2,3}$) applied across pairs of legs and the top (grounded) contact, allow the in-plane stress in the QD membrane to be controlled. An additional ring contact on top of the membrane enables electric-field control across the QDs. **(e)**. Same as in (b) for two remote QDs embedded in the device shown in (d).

**Figure 2. (a)**. Mismatch between $\theta_+$ and $\phi_\varepsilon$ in the condition $\phi_\varepsilon = \phi_\varepsilon^*$. The blue (red) line indicates the result for InAs (GaAs) QDs. The inset shows a sketch of a QD featuring $C_1$ symmetry and the stress applied to the QD in terms of $S_1$, $S_2$ and $\phi$. **(b)**. FSS against the X energy-shift ($E_0 - E_X$) for a InAs QD with $\theta_+ = 16.2°$, with $E_0$ the X energy at 0 stress. The colored lines correspond to $S_1$ applied at $\phi = 23°$ (purple line), $\phi = 0°$ (red line), and $\phi = 11°$ (blue line) and with $S_2 = 0$. The black lines correspond to $\phi_\varepsilon = \phi_\varepsilon^* = 18.5°$ and for $S_2 = 0$, 0.5 and 1800 MPa, from left to right respectively. The values of the $\phi_\varepsilon^*$ $\Delta p^*$ and $\bar{p}$ at which $s=0$ are also indicated.

**Figure 3. (a)**. FEM simulations of the $S_1$ for two sets of voltages: $(V_1,V_2,V_3)=(0,600,0)$V (top) and $(V_1,V_2,V_3)=(600,0,0)$V (bottom). The insets show zoom of the central regions and the angle formed by $S_1$ with respect to the [110] direction. We point out that a major stress of 1.8 GPa along the [100] direction corresponds to a major strain of ~1.5%. **(b)**. X energy against $\phi$ and $\Delta p$ for $(V_1,V_2,V_3)$ in the range ±600 V. A cylindrical coordinate system has been used.

ACKNOWLEDGMENT

We thank J. S. Wildmann, T. Lettner for help. The work was supported financially by the European Union Seventh Framework Programme 209 (FP7/2007-2013) under Grant Agreement No. 601126 210 (HANAS), and the AWS Austria Wirtschaftsservice, PRIZE Programme, under Grant No. P1308457.

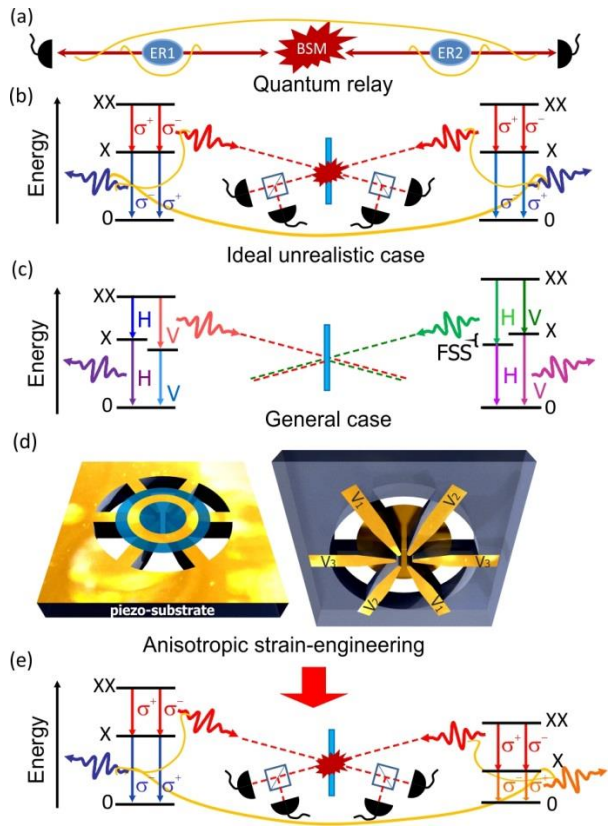

Figure 1 of 3

by R. Trotta *et al*.

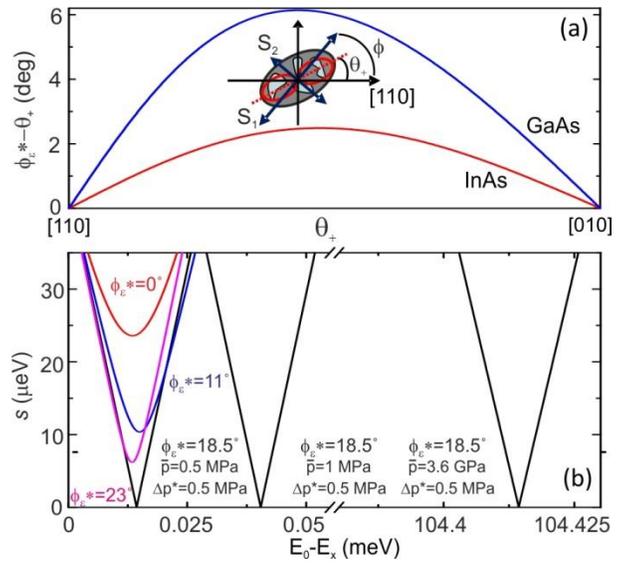

Figure 2 of 3

by R. Trotta *et al*.

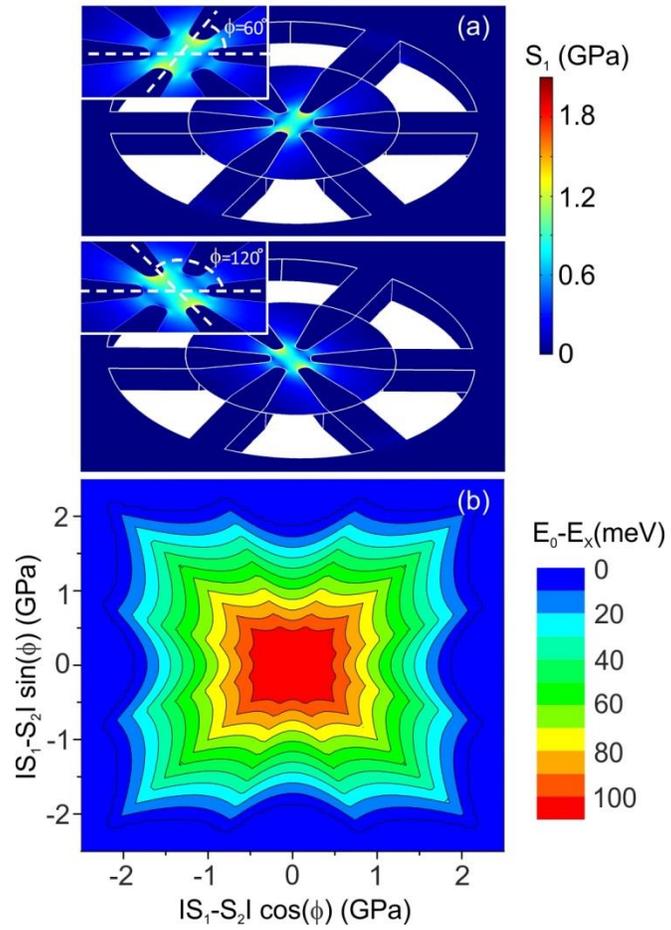

Figure 3 of 3
by R. Trotta *et al*.